# A Finite Element Analysis Model for Magnetomotive Ultrasound Elastometry Magnet Design with Experimental Validation

**Jacquelline Nyakunu[1], Christopher T. Piatnichouk[1], Henry C. Russell[1], Niels J. van Duijnhoven[1], and Benjamin E. Levy[1]**

[1] Department of Physics, Davidson College, Davidson, NC 28035, United States of America

E-mail: belevy1@davidson.edu

## Abstract

*Objective*. Magnetomotive ultrasound (MMUS) using magnetic nanoparticle contrast agents has shown promise for thrombosis imaging and quantitative elastometry via magnetomotive resonant acoustic spectroscopy (MRAS). Young's modulus measurements of smaller, stiffer thrombi require an MRAS system capable of generating forces at higher temporal frequencies. Solenoids with fewer turns, and thus less inductance, could improve high frequency performance, but the reduced force may compromise results. In this work, a computational model capable of assessing the effectiveness of MRAS elastometry magnet configurations is presented and validated. *Approach*. Finite element analysis (FEA) was used to model the force and inductance of MRAS systems. The simulations incorporated both solenoid electromagnets and permanent magnets in three-dimensional steady-state, frequency domain, and time domain studies. *Main results*. The model successfully predicted that a configuration in which permanent magnets were added to an existing MRAS system could be used to increase the force supplied. Accordingly, the displacement measured in a magnetically labeled validation phantom increased by a factor of $2.2 \pm 0.3$ when the force was predicted to increase by a factor of $2.2 \pm 0.2$. The model additionally identified a new solenoid configuration consisting of four smaller coils capable of providing sufficient force at higher driving frequencies. *Significance*. These results indicate two methods by which MRAS systems could be designed to deliver higher frequency magnetic forces without the need for experimental trial and error. Either the number of turns within each solenoid could be reduced while permanent magnets are added at precise locations, or a larger number of smaller solenoids could be used. These findings overcome a key challenge toward the goal of MMUS thrombosis elastometry, and simulation files are provided online for broader experimentation.

## 1. Introduction

Ultrasound elastometry has clinical uses across multiple organs in adult and pediatric patients[1,2], including the diagnosis, staging, and treatment monitoring of cancer and liver disease[3,4]. In deep vein thrombosis (DVT), elastometry can provide thrombus age and composition data, and the technique shows potential for aiding in treatment decisions[5,6]. To meet the need for targeted elastometry, an emerging technique called Magnetomotive Ultrasound (MMUS) is currently under development toward quantitative elastometry applications. First introduced in 2006[7] and recently demonstrated in human tissue[8], MMUS is a contrast-enhanced imaging technique wherein subresolution magnetic nanoparticle (MNP) contrast agents are driven by a time-varying magnetic field, and the resulting tissue motion is tracked via ultrasound[9]. MMUS-based elastometry was initially investigated for colorectal cancer screening in 2015[10]. Subsequent work has shown that shear waves produced in the imaging process can be used to reconstruct quantitative elastograms[11], and to monitor the change in tissue elastic properties subject to magnetic hyperthermia[12]. Recently, MMUS has been investigated for its potential in thrombosis imaging[13,14]. MMUS-based resonant acoustic spectroscopy (MRAS) has been shown to be capable of quantitative elastometry in thrombus mimicking structures[15], despite challenges in magnet design that currently limit applicability. In this work, a finite element analysis (FEA) model for the design of MMUS magnet systems is presented and validated. The work may be useful in a wide range of clinical and preclinical applications such as magnetic drug targeting and magnetic hyperthermia, where well-designed MMUS systems could improve on MRI and magnetic particle imaging (MPI)-based localization[16]. However, in this study, particular focus is paid to meeting the requirements for thrombus elastometry.

Significant progress toward quantitative MMUS-based elastometry has been reported in recent years, but key challenges remain. Elastometry is commonly accomplished by applying a known stress and measuring the resulting strain, but in MMUS, concentrations and spatial distributions of MNPs, and thus applied forces, are generally unknown. To overcome this, Thapa[17], and Fink [18] used inverse methods and FEA, respectively, to reconstruct body force distributions, but these approaches rely on homogenous MNP distributions and tissue properties which may not hold in all cases. Furthermore, Sjöstrand created an MNP-labeled microbubble contrast agent that could be visualized directly via conventional ultrasound[19], although it is unclear whether these relatively large bubbles would possess all the same labeling capabilities as nanoparticles. Avoiding the need for known particle distributions, Lin recently demonstrated that the speed of shear waves resulting from magnetically induced MNP motion can be measured quantitatively *in vivo*[11]. These shear waves may also be visualized qualitatively via Eulerian motion magnification[20]. While exciting, this shear wave elastography approach is likely better suited to assessing tissue properties outside of a small, labeled region rather than measuring the interior of a labeled structure such as a thrombus or a tumor. Conversely, MRAS has been demonstrated to be capable of quantitative elastometry *within* magnetically labeled model blood clots[15]. In this approach, the resonance frequencies of magnetically labeled regions are determined via MMUS, and geometric data is collected with conventional ultrasound. These measurements are then input into a model that uses FEA to determine the Young's modulus.

To make MRAS measurements, a chirped sinusoidal magnetic force capable of output up to the highest expected resonance frequency is needed. Resonance frequencies tend to increase for smaller, stiffer structures, so a future MRAS system designed for thrombus elastometry should be capable of matching the resonance frequencies of low volume (0.1 ml [21,22]), high Young's modulus (80 kPa[23]) clots. Using these parameters and the FEA model presented in previous work[15], a semiellipsoid designed to match the shape of a thrombus[24] was simulated by the authors, and the resonance frequency was measured to be 188 Hz. Thus, magnets capable of supplying a 200 Hz driving force should be sufficient in most cases. However, the current MRAS system is limited to approximately 50 Hz by the large inductance of its solenoid electromagnets[15]. Decreasing the number of turns in each solenoid would decrease the inductance, but it could also have a deleterious effect on the system's performance due to a reduction in magnetic force[25]. One solution involves either placing a single solenoid directly below the sample and opposite the transducer[9], or surrounding the sample[26]. Although this would reduce the necessary range over which the force must act, and thus would allow for smaller, lower-inductance coils, the current "open-air" arrangement with solenoids located beside the transducer is more clinically relevant as it can accommodate arbitrarily large samples[14]. Determining alternate magnet orientations and geometries may allow for improved high frequency performance.

As first reported by Wang[27], the addition of a permanent magnet stuck to the core of an MMUS electromagnet may pre-magnetize MNPs in the imaging area and increase the amplitude of the driving force without the need for additional solenoid turns. Reniaud subsequently demonstrated that a rotating permanent magnet coupled with a spatially homogeneous, temporally modulated field from a Helmholtz coil could increase magnetically induced motion by about 40% [28,29]. These advances raise the prospect of a future MMUS system designed for high frequencies through the combination of smaller, lower-inductance solenoids and permanent magnets. However, the optimal position for these permanent magnets is far from obvious as it requires careful accounting for multiple three-dimensional, inhomogeneous fields. Toward the goal of long-range, high frequency magnetic forces, an FEA simulation capable of predicting the performance of alternate MMUS magnet arrangements is presented and validated. The model is initially used to predict an effective placement for permanent magnets to enhance an existing MMUS system. Then, to address practical and safety-related concerns associated with the presence of a permanent magnetic field, the model is extended to assess a new arrangement that could improve high frequency performance without permanent magnets.

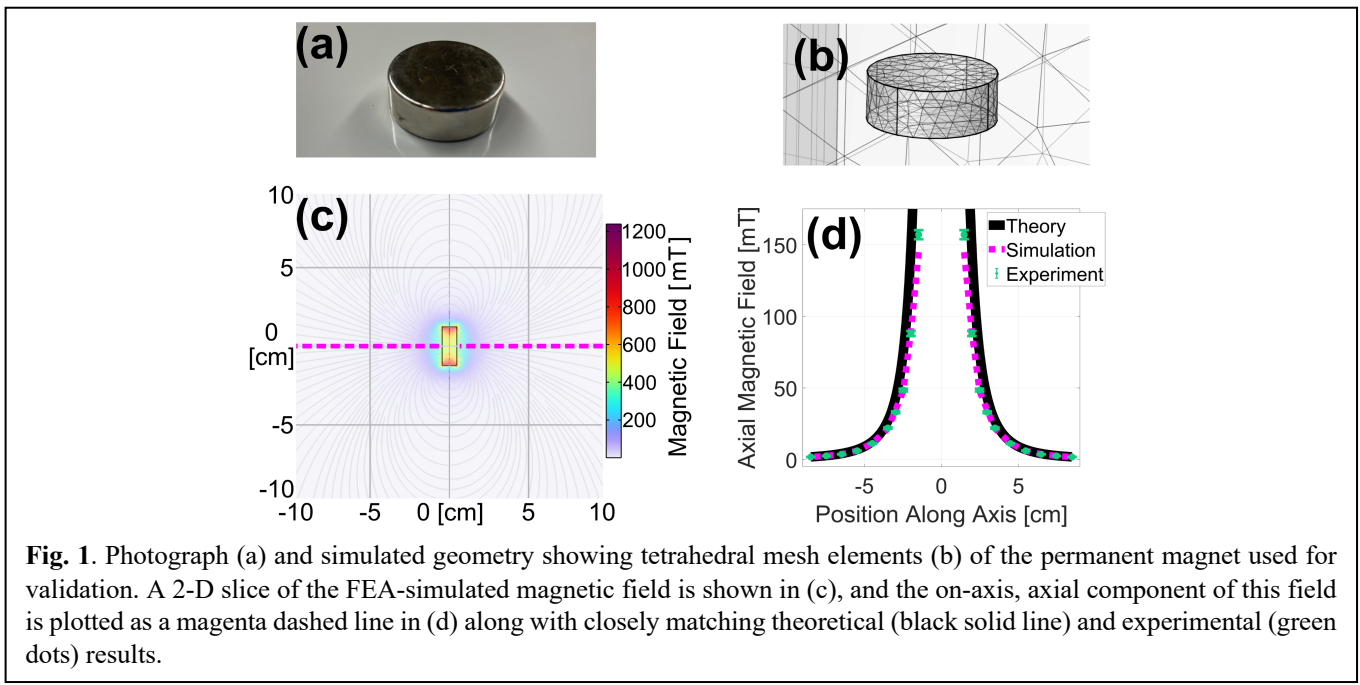

**Fig. 1**. Photograph (a) and simulated geometry showing tetrahedral mesh elements (b) of the permanent magnet used for validation. A 2-D slice of the FEA-simulated magnetic field is shown in (c), and the on-axis, axial component of this field is plotted as a magenta dashed line in (d) along with closely matching theoretical (black solid line) and experimental (green dots) results.

## 2. Methods

All simulations were conducted using the "Magnetic Fields (mf)" interface within the AC/DC module of COMSOL Multiphysics® v. 6.1 (COMSOL AB, Stockholm, Sweden). Each simulation file is available for download online[30].

### 2.1 Validation of FEA Models

Before drawing conclusions, it was necessary to confirm whether the finite element methods to be used could match relevant experimental and theoretical results.

#### 2.1.1 Permanent Magnet

Starting with permanent magnets, a study was conducted comparing simulated, experimental, and theoretical data. As shown in Fig. 1a, a 1.0 in diameter × 3/8 in tall cylindrical N52 grade neodymium iron boron (NdFeB) permanent magnet (DX06-N52, K&J Magnetics, Inc, Pipersville, PA) with a dipole moment $\vec{m} = 5.68$ Am$^2$ was used for this study. The theoretical magnetic field strength was estimated utilizing the dipole approximation[31],

$$\vec{B}(r) = \frac{\mu_0}{4\pi r^3} \left( -\vec{m} + 3(\vec{m} \cdot \hat{r})\hat{r} \right), \quad (1)$$

where $r$ is the radial distance from the dipole. When $\vec{m}$ is aligned with the $z$-axis, the axial component of the magnetic field is given by

$$B_z = \frac{\mu_0 m}{2\pi |z|^3}. \quad (2)$$

Experimental measurements were obtained using the transverse probe of a handheld gauss meter (F.W. BELL

Model 4048, Milwaukie, OR, USA). An FEA simulation was then created with a magnet of the same dimensions, a recoil permeability of 1.1, a remanent flux density norm of 1.4 T, and an electrical conductivity of 0.71 MS/m. The magnet was enclosed by a cubical air domain approximately ten times the size of the magnet. The air was set to zero electrical conductivity with relative permittivity and permeability values of 1.0. As illustrated in Fig. 1b, a tetrahedral mesh was applied across all domains with an average element volume of 420 mm$^3$ in the air, and 5.2 mm$^3$ in the magnet where more precision was necessary. Ampère's law was solved in a stationary study on all domains, and the exterior boundaries of the air domain were set to be magnetic insulators. The top and bottom surfaces of the cylindrical magnet were designated as the north and south magnetic poles, respectively.

Figure 1c shows magnetic field lines and field strength as predicted from the stationary study. As shown in Fig. 1d, the theoretical (Eq. 2), simulated, and experimental results with manufacturer specified uncertainties compare favorably. The axial magnetic field strength decreases as distance increases from the magnet. The slight deviation in theoretical results close to the magnet is expected as the dipole approximation breaks down in the near limit.

#### 2.1.2 Solenoid

To validate the model's performance with solenoids, a similar study was conducted comparing FEA simulation results against theoretical and experimental data. The solenoid shown in Fig. 2a had an 8.9 mm inner diameter, a length of $L = 45$ mm, and $N = 38$ turns of tightly packed 18 AWG enameled copper wire with current $I = 1.00$ A.

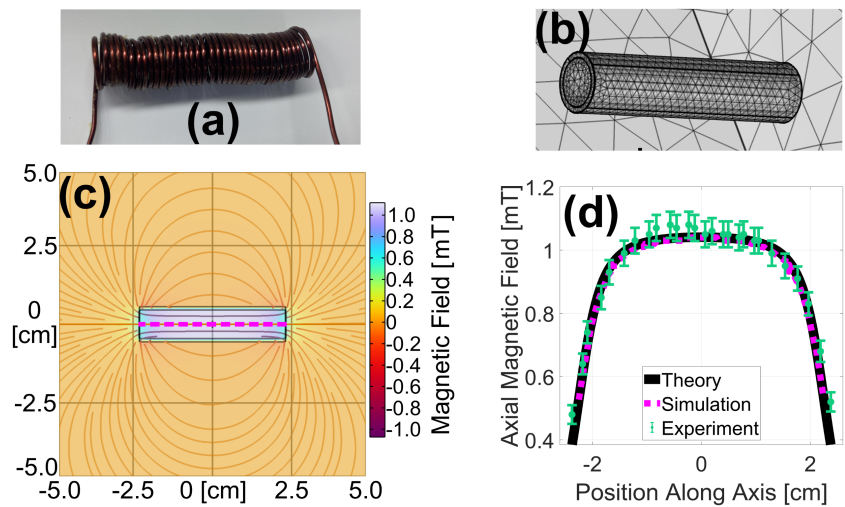

**Fig. 2.** Photograph (a) and simulated geometry showing tetrahedral mesh elements (b) of the solenoid used for validation. A 2-D slice of the FEA-simulated magnetic field is shown in (c), and the on-axis, axial component of this field is plotted as a magenta dashed line in (d) along with closely matching theoretical (black solid line) and experimental (green dots) results.

Experimentally, a gaussmeter was inserted into the solenoid to measure the B-field along the coil's *z*-axis. Theoretically, the *z*-component of the on-axis magnetic field from a finite continuous solenoid of radius $R$ can be expressed as[32],

$$B_z = \frac{\mu_o N I}{2} \left( \frac{\frac{L}{2} - z}{L\sqrt{R^2 + \left(\frac{L}{2} - z\right)^2}} + \frac{\frac{L}{2} + z}{L\sqrt{R^2 + \left(\frac{L}{2} + z\right)^2}} \right). \quad (3)$$

A solenoid of the same dimensions was simulated as a "homogenized multiturn coil," an approximation that reduces computation time, but neglects the skin effect and capacitive coupling. The coil depicted in Fig. 2b was set to have the properties of copper wire with a relative permeability and permittivity of 1.0, and an electrical conductivity of 60 MS/m. An air domain similar to that described in Sec. 2.1.1 was used and the same physics was applied. The average mesh element volume in the solenoid was 0.76 mm³. A "coil geometry analysis" step preceded the stationary study. The resulting magnetic field lines and axial field strength are plotted in Fig. 2c. The matching theoretical (Eq 3), experimental, and simulation results shown in Fig. 2d lend further credibility to this FEA simulation.

### 2.1.3 Existing MMUS System

To further validate the model's relevance for MMUS, a simulation that matched an existing open-air MMUS apparatus described in previous work[14,33] and shown in Fig. 3a was created. The existing system consisted of two water-jacketed solenoid electromagnets flanking an Ultrasonics L14/5-38 transducer (Analogic Corporation, Peabody, MA, USA) with 1100 turns apiece of 18 AWG copper wire, and grain oriented electrical steel (GOES) cores. The magnets were driven by a square root sinusoidal current excitation with a peak current of 12.6 A. The coordinate system shown in Fig. 3b was used throughout this work. Sound emanates from the transducer along the axial direction, *z*. The lateral direction *x* signifies the perpendicular coordinate in the imaging plane, and the elevational direction *y* denotes the out of plane coordinate. In previous work, this configuration was used to measure the magnetic force imparted on a 0.5 mm-diameter (AISI 52100) chrome steel ball positioned at various locations across the imaging area[34]. Those results are illustrated in Fig. 3c.

The simulation depicted in Fig. 3b was designed to match the materials and dimensions of the existing apparatus. The GOES cores were simulated with an electrical conductivity of 2.08 MS/m, a relative permittivity of 1.0, and a relative permeability of 4000. Homogenized multiturn copper coils were modeled as in Sec. 2.1.2 to surround the cores. A 0.5 mm-diameter chrome steel ball was placed 10 mm axially below the cores, was laterally and elevationally centered, and had relative permeability, permittivity, and conductivity values of 4.0[35], 1.0, and 1.5 MS/m, respectively. The apparatus was enclosed by an air domain similar to that employed in Sec. 2.1.1 and a tetrahedral mesh was applied across all domains. The custom mesh ranged from an average element volume of 980 mm³ in the air to 7,700 µm³ in the chrome steel ball where greater precision was needed. To create a smoother transition, elements in the 1.0 cm radius surrounding the ball were assigned an intermediate value of 0.24 mm³. The electromagnets had an average element volume of 31 mm³. The same physics discussed in Secs. 2.1.1 and 2.1.2 was applied, with an additional force calculation added to the ball. The existing system operates in the static limit, so a steady-state study with current corresponding to the peak experimental value was performed. Forces were computed at each point shown in Fig. 3d, utilizing cubic interpolation for the colormap. The experimentally determined force map depicted in Fig. 3c closely aligns with the simulation results. Excluding the bottom 1.0 cm axially, the average error was 31%. In general, the simulation slightly overestimated the force at larger axial depths. Excluding the bottom 1.0 cm axially, the average error fell to 16%, and the error at the point centered 1.0 cm below the transducer was <5%. This ball location was therefore utilized for subsequent studies.

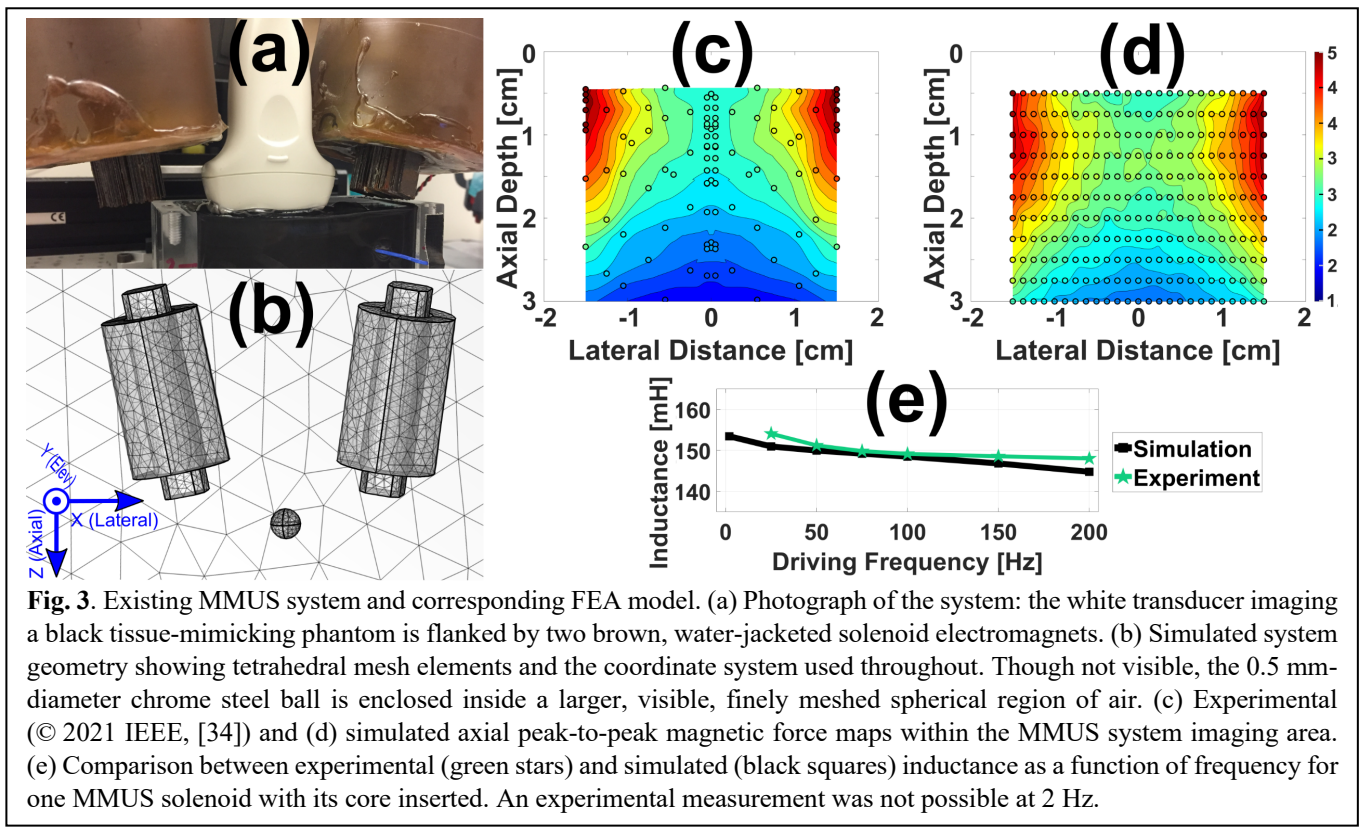

**Fig. 3.** Existing MMUS system and corresponding FEA model. (a) Photograph of the system: the white transducer imaging a black tissue-mimicking phantom is flanked by two brown, water-jacketed solenoid electromagnets. (b) Simulated system geometry showing tetrahedral mesh elements and the coordinate system used throughout. Though not visible, the 0.5 mm-diameter chrome steel ball is enclosed inside a larger, visible, finely meshed spherical region of air. (c) Experimental (© 2021 IEEE, [34]) and (d) simulated axial peak-to-peak magnetic force maps within the MMUS system imaging area. (e) Comparison between experimental (green stars) and simulated (black squares) inductance as a function of frequency for one MMUS solenoid with its core inserted. An experimental measurement was not possible at 2 Hz.

Finally, to verify FEA coil inductance predictions, a frequency domain study ranging from 2.0 Hz to 200 Hz measured the inductance of the coil. Experimentally, the inductance was determined by measuring the resonance of an RLC circuit. The average error between the experiment and simulation lines shown in Fig. 3e was 1.2 ± 0.8% for frequencies ≥ 25 Hz.

## 2.2 Augmenting Existing MMUS System with Permanent Magnets

Having validated the performance of the presented FEA approach under relevant limiting cases and experimental conditions, the potential for permanent magnets augmenting the existing MMUS solenoids to yield a greater force was then explored.

### 2.2.1 Theoretical Underpinning

To avoid unnecessary computation, potential options were first narrowed based on conclusions drawn from the governing physics. Assuming remnant magnetization is low, the force on a particle of volume $V$ and volume magnetic susceptibility $\chi$ produced by a magnetic field $\vec{B}$ is[9],

$$\vec{F} = \frac{1}{2}\frac{V\chi}{\mu_0}\vec{\nabla}|\vec{B}^2| \approx \frac{V\chi}{\mu_0}(\vec{B}\cdot\vec{\nabla})\vec{B}, \qquad (4)$$

where the approximation holds when $\vec{\nabla}\times\vec{B}\approx 0$. This is a reasonable assumption because magnet excitation frequencies were kept relatively low in this study, and there was no current density in the imaging area. Thus, the axial component of this force can be expressed by the proportionality

$$F_z \propto B_x\frac{\partial B_z}{\partial x} + B_y\frac{\partial B_z}{\partial y} + B_z\frac{\partial B_z}{\partial z}, \qquad (5)$$

where an increase in any one of the three terms may increase MMUS signal. It was hypothesized that augmenting the existing MMUS system with cylindrical permanent magnets oriented with their axes in the lateral direction would produce both a large $B_x$ and an appreciable $\partial B_z/\partial x$.

### 2.2.2 FEA Simulation

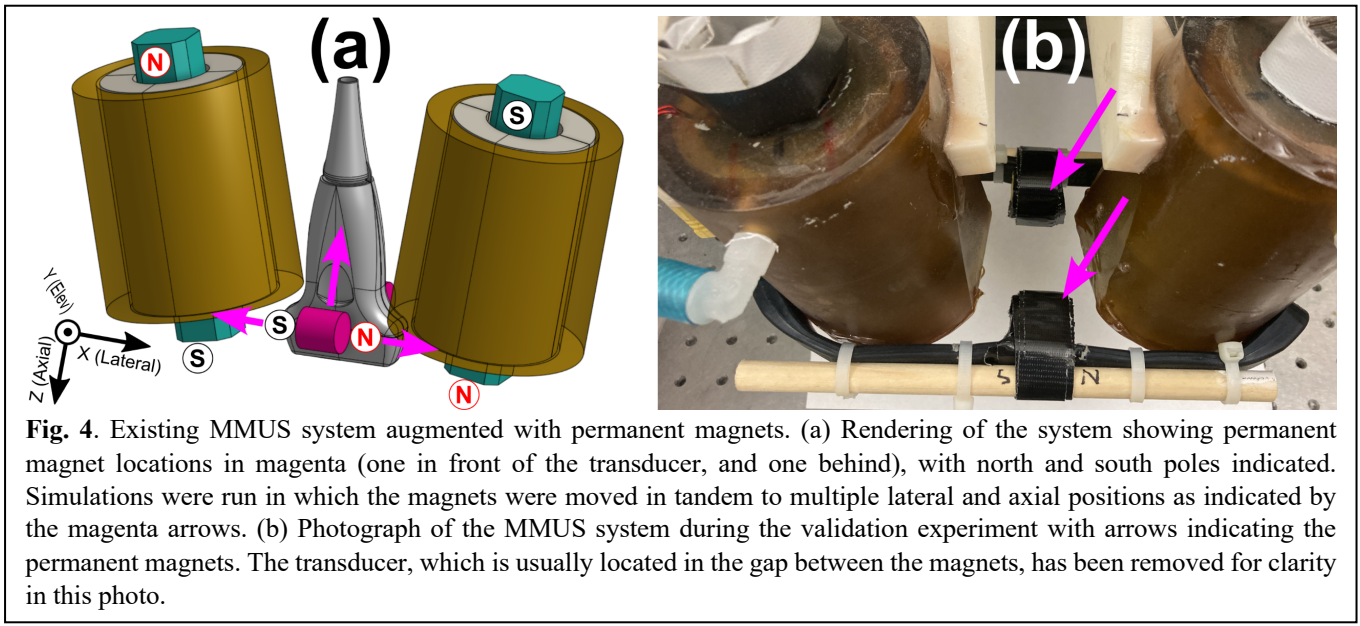

**Fig. 4**. Existing MMUS system augmented with permanent magnets. (a) Rendering of the system showing permanent magnet locations in magenta (one in front of the transducer, and one behind), with north and south poles indicated. Simulations were run in which the magnets were moved in tandem to multiple lateral and axial positions as indicated by the magenta arrows. (b) Photograph of the MMUS system during the validation experiment with arrows indicating the permanent magnets. The transducer, which is usually located in the gap between the magnets, has been removed for clarity in this photo.

The FEA model from Sec. 2.1.3 was adapted to include two identical cylindrical N52 grade NdFeB permanent magnets as shown in magenta in Fig. 4a. Each magnet could be set to an equal diameter and height of 1.0 in, 3/4 in, or 5/8 in, in order to match commercially available offerings. The center of each cylinder was fixed 2.5 cm away elevationally from the transducer center, and could be translated axially and laterally as indicated by the magenta arrows in Fig. 4a. Besides the permanent magnets, which were assigned parameters from Sec. 2.1.1 and an average tetrahedral mesh volume of 33 mm³, the same material, geometry, physics, and meshing properties were employed as in Sec. 2.1.3. However, instead of relying on steady-state assumptions, time-dependent studies were performed. Current through the solenoids varied square-root sinusoidally with a 12.6 A peak and a 2.0 Hz frequency. To overcome a technical limitation of COMSOL Multiphysics® v. 6.1 requiring a current return path in time dependent studies, the air domain was given an electrical conductivity of 1.0 S/m. This did not significantly impact the simulated results as the air conductivity is much less than that of a conductor.

Exploratory studies found that when both permanent magnets were positioned with their north poles to the right as shown in Fig. 4a, a substantial increase in axial force on the steel ball discussed in Sec. 2.1.3 was possible. To determine an effective placement, the parallel permanent magnets were moved in tandem to various axial and lateral positions relative to the transducer. For each of the three magnet sizes, they were positioned at lateral distances ranging from -2.5 to 2.5 cm in steps of 0.5 cm and axial distances of -12 to -2.0 cm (i.e., above the transducer) in steps of 1.0 cm. For each magnet position, force was recorded as a function of time for 0.5 s at a sampling frequency of 200 samples/s, and the peak-to-peak force was recorded.

### 2.2.3 Experimental Validation

To verify whether the addition of permanent magnets in FEA-predicted locations could lead to increased magnetic forces, a validation experiment was conducted. First, a tissue-mimicking phantom with a Young's modulus of 5.0 kPa was constructed following a previously published recipe[33]. The phantom was primarily composed of gelatin and water with added graphite nanopowder and alcohol to act as an acoustic scattering agent and a speed of sound adjustment, respectively. A 1.0 ml cubical inclusion composed of the same material was labeled with iron oxide nanopowder at a concentration of 5.0 mg Fe/ml. Next, two cylindrical 0.75 in (height and diameter) N52-grade permanent magnets (DCC-N52, K&J Magnetics, Inc, Pipersville, PA) were affixed to the existing MMUS solenoids. As shown in Fig. 4b, rubber straps, zip ties, wooden dowels, and tape were used for mounting. To match the simulation geometry shown in Fig. 4a as closely as possible, both magnets were mounted with their axes parallel to the $x$ direction and their north poles on the right. The magnets were centered laterally, placed at $z = -2.2 \pm 0.1$ cm axially, and elevationally separated from the imaging plane by $3.0 \pm 0.1$ cm.

Previously reported imaging and image processing protocols were followed[13]. Briefly, to form an image two 7.5 s stacks of beamformed RF ultrasound data were collected, one with the solenoids producing a 2.0 Hz square-root sinusoidal force, and the next with the solenoids turned off for background subtraction. These data are available for download[30]. A frequency and phase-locking algorithm was employed to create a quantitative "MMUS image" displaying the magnetically induced displacement amplitude at all locations. Three MMUS images were obtained with the permanent magnets present, and three more were obtained with the permanent magnets removed. Datasets were collected in an alternating fashion to rule out the possibility of time-related effects. Subsequently, a version of the simulation presented in Sec. 2.2.2 was run with permanent magnet position parameters exactly matching the experiment described in the current section.

### 2.3 Four Smaller Solenoids

In order to allow for sufficient force at frequencies up to 200 Hz, a new magnet configuration consisting of four smaller solenoids was proposed, and the FEA model was used to refine the proposal and predict its efficacy. Maintaining high AC current at higher frequencies can be achieved by lowering the inductance. Because an increasing number of turns in a solenoid leads to only a linear increase in the magnetic field while inductance scales quadratically, keeping the number of turns low is advantageous. Therefore, it was hypothesized that four solenoids each with fewer turns than the existing MMUS magnets should be able to achieve the same magnetic field strength in the imaging area while allowing for the higher frequencies necessary in MRAS.

#### 2.3.1 Solenoid Geometry

Figure 5 presents the proposed magnet configuration. The solenoids contained coils of length $L$ and iron cores of length $l$. The length of the coils was varied, and the length of cores was set to be $l = L + 5.0$ cm. The diameter $d$ of the iron cores was fixed at 2.0 cm while the diameter $D$ of the coils was fixed at 4.0 cm. All four coils could be tilted an angle $\theta$ from the vertical. The coil density was kept the same as in the existing MMUS system, which was 1100 turns for a cross-sectional area of 15.75 cm², or 69.84 turns/cm². Therefore, the number of turns can be calculated via

$$N = L \times \left( \frac{D - d}{2} \right) \times 69.84 \text{ turns/cm}^2. \qquad (6)$$

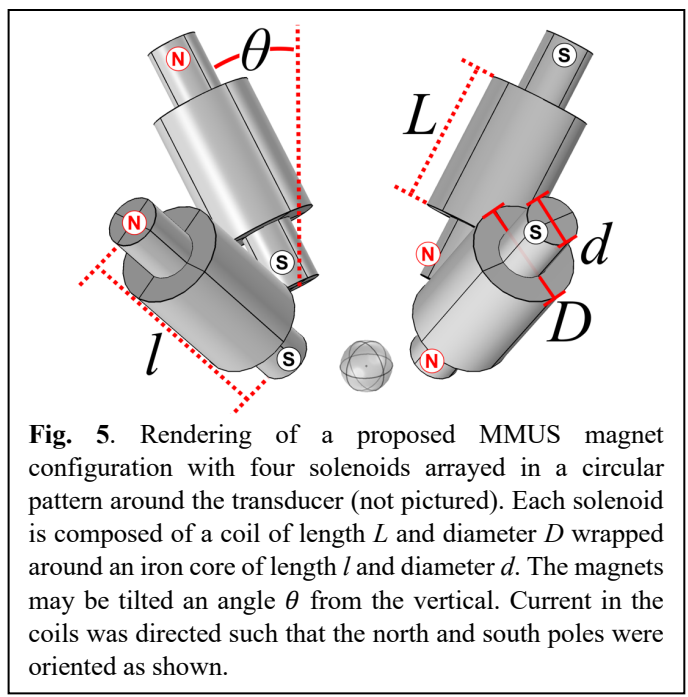

**Fig. 5**. Rendering of a proposed MMUS magnet configuration with four solenoids arrayed in a circular pattern around the transducer (not pictured). Each solenoid is composed of a coil of length $L$ and diameter $D$ wrapped around an iron core of length $l$ and diameter $d$. The magnets may be tilted an angle $\theta$ from the vertical. Current in the coils was directed such that the north and south poles were oriented as shown.

In this geometry, two of the solenoids would flank the L14/5-38 transducer laterally, spaced for minimum clearance, while the other two would be separated by the same distance elevationally. Current was directed such that opposite solenoids had antiparallel poles as shown in Fig. 5. The material and mesh properties described in Sec. 2.1.3 were once again employed for this simulation with the exception of the electromagnet mesh, which had a slightly smaller average element volume of 18 mm³. Furthermore, this simulation used the same physics described in Sec. 2.2.2.

#### 2.3.2 Simulation Details

The goal of this simulation was to determine a solenoid configuration with minimal inductance that could match the force of the existing 2 Hz system up to a 200 Hz driving frequency. First, the same square-root sinusoidal current waveform from Sec. 2.2.2 was used and force was simulated in the time domain for five cycles at 20 samples/cycle with driving frequencies ranging from 2.0 to 200 Hz. The peak-to-peak force was determined from the fifth cycle to minimize the impact of any potential transients. Each simulation was repeated for coil lengths of $L = 25, 50, 55, 60, 75,$ and $100$ mm. Next, inductance was simulated in the frequency domain for a single solenoid in the presence of the other three solenoids over the same parameter space. Coil length was subsequently fixed at 55 mm, and time- and frequency-domain sweeps were repeated for tilt angles of $\theta = 0°$ to $70°$ in steps of $10°$.

### 3. Results

Figure 6 displays the FEA model prediction for the peak-to-peak axial 2.0 Hz magnetic force imparted on a hypothetical 0.5 mm diameter chrome steel ball by the existing open-air MMUS system[14] with two additional NdFeB permanent magnets. The ball was positioned 1.0 cm below the face of the ultrasound transducer as discussed in Sec. 2.1.3 and centered laterally. In Fig. 6a, the axial positions of the two permanent magnets were fixed at $z = -2.0$ cm (3.0 cm above the chrome steel ball), and the lateral positions were fixed at $y = \pm 2.5$ cm, with force plotted as a function of lateral magnet position. In Fig. 6b, the lateral magnet positions were instead fixed at $x = 0$ cm (centered), with force plotted as a function of axial position. Note that $z = 0$ is defined as the transducer face, with larger negative numbers indicating movement upward, away from the imaging area. Three different pairs of cylindrical magnets were tested, each with equal diameter and height as indicated in the legend. Peak-to-peak force is maximized in each case when the magnets are laterally centered and axially as close to the imaging area as possible. Reassuringly, the force approaches 3.2 μN when the permanent magnets are moved far from the imaging area. As shown in Fig. 3c, this is the known force applied by the existing MMUS system to chrome steel balls located in the same position.

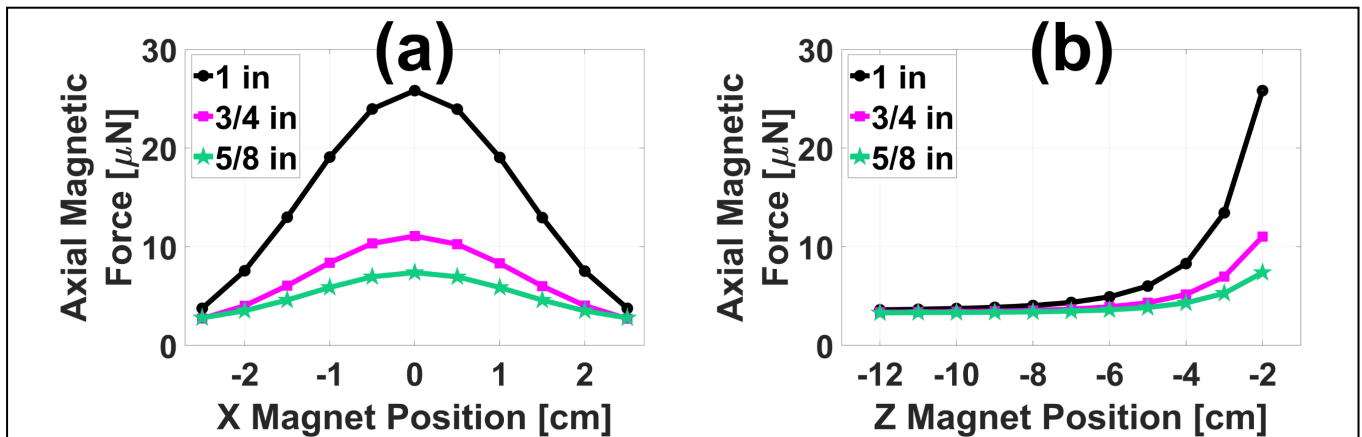

**Fig. 6**. FEA permanent magnet study results: Magnetic force on a 0.5 mm diameter chrome steel ball subject to the addition of two permanent magnets augmenting the existing MMUS system as shown in Fig. 4(a). Peak-to-peak axial force is plotted as a function of lateral (a) and axial (b) magnet position for three pairs of permanent magnets. The 1 in (diameter and length) pair is shown with black dots, the 3/4 in pair is shown with magenta squares, and the 5/8 in pair is shown with green stars. In (a), the axial position of the magnets is fixed at $z = -2.0$ cm, and in (b), the lateral position is fixed at $x = 0$ cm.

Figure 7 displays validation results for the model presented above. The FEA-predicted 2.0 Hz force as a function of time for the existing system on the hypothetical ball is shown as a black dashed line in Fig. 7a. With the addition of two 3/4 in NdFeB magnets positioned at $x = 0$ cm, $y = \pm 3.0$ cm, and $z = -2.2$ cm, the predicted peak force increases by a factor of $2.2 \pm 0.2$ to the solid blue line. Figure 7b shows a standard ultrasound brightness-mode (B-mode) image of the tissue mimicking phantom with a 1.0 ml cubic, magnetically-labeled inclusion. The three bright white spots are fiducial markers, placed as described in previous work[13] to allow the precise location of the inclusion to be selected. Figures 7c and 7d are representative MMUS images of the phantom without and with the 3/4 in permanent magnets inserted, respectively. These data were collected with a 2.0 Hz temporal magnet modulation frequency to match the simulation. The inclusion location is outlined with a red square. Averaging over three images in each case, the magnetically induced displacement amplitude (MMUS signal) in the inclusion area was measured to be $17 \pm 2$ nm without the permanent magnets, and $38.6 \pm 0.8$ nm with the permanent magnets. This increase by a factor of $2.2 \pm 0.3$ agrees with the FEA-predicted factor of $2.2 \pm 0.2$, bolstering the results in Fig. 6.

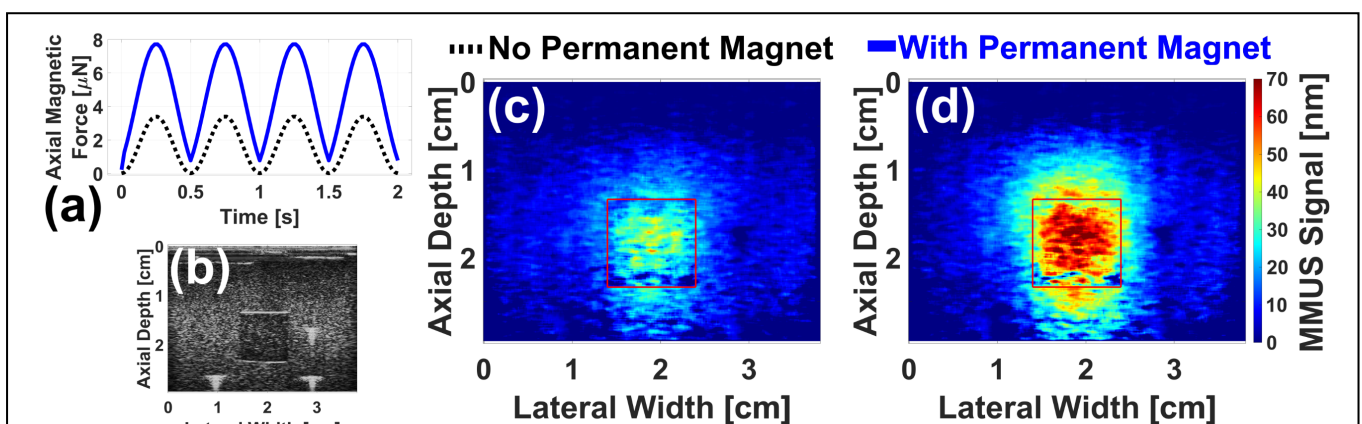

**Fig. 7**. Permanent magnet experimental validation results: (a) FEA-predicted axial magnetic force on a 0.5 mm diameter chrome steel ball is plotted as a function of time without permanent magnets (black dashed line) and with two 3/4 in permanent magnets placed at $x = 0$ cm and $z = -2.2$ cm (blue solid line). (b) Standard ultrasound B-mode image of the gelatin tissue mimicking phantom used for validation. Representative MMUS images with 2.0 Hz modulation frequencies displaying the magnetically induced displacement amplitude within the phantom are presented for the case without (c) and with (d) the permanent magnets. In each image, a square is drawn around the magnetically labeled region.

While investigation of the existing open-air system was limited to a 2.0 Hz driving frequency for consistency with the limitations outlined in Sec. 1, the hypothesized system of four smaller solenoids discussed in Sec. 2.3 was simulated through a range of higher frequencies to test its performance. Figure 8 presents these results. As seen in Fig. 8a, increasing coil length, and the resulting increase in turns via Eq. 6, lead to higher peak-to-peak magnetic forces on a hypothetical chrome steel ball when the tilt angle was fixed at $\theta = 0°$. This trend held across all frequencies tested, although overall force was lower at high frequencies. This is likely due to increased inductive reactance. Longer coils with more turns also resulted in higher single-coil inductance values across all frequencies, as shown in Fig. 8b. However, the simulation also predicts a significant frequency dependence, especially at high coil lengths. This does not contradict the previous finding as inductive reactance is proportional not only to inductance, but also to frequency.

According to Figs. 8a and 8b, 5.5 cm-long coils (384 turns each per Eq. 6) represent the minimum length for which the peak-to-peak force across all frequencies meets or exceeds the 3.2 µN supplied by the existing system at 2.0 Hz. Furthermore, the inductance of each coil is predicted to be less than 15% of that shown in Fig. 3e for an existing MMUS solenoid. The tilt angle of the four 5.5 cm-long solenoids was varied from $\theta = 0°$ to $\theta = 70°$, and results for force and displacement are shown in Fig. 8c and 8d, respectively. The simulation predicts that a 0° tilt angle will produce the largest force without changing inductance substantially.

## 4. Discussion

The FEA model presented in this work predicted that two potential MMUS magnet configurations could improve the performance of an MRAS elastometry system. First, existing MMUS solenoids were augmented with two additional permanent magnets, and the model was used to determine magnet orientations and locations that would lead to an increased time-varying force. Large permanent magnets positioned close to the imaging area with their axes aligned in the lateral direction were predicted to be capable of producing the largest increase, with a 1-in (length and diameter) pair increasing the force by a factor of more than 8 as shown in Fig. 6a. The model was validated experimentally for an intermediate configuration with 3/4-in permanent magnets positioned slightly farther from the imaging area. The FEA model predicted a factor of $2.2 \pm 0.2$ increase in force, while experimentally the displacement that an MNP-labeled gelatin inclusion increased by a factor of $2.2 \pm 0.3$. These two results agree as expected under linear elastic assumptions and affirm the model's utility for designing future MMUS magnets. Uncertainty in the FEA result accounts for imprecise knowledge of the exact permanent magnet locations, while uncertainty in the experimental results is approximated via repeated measurements. Furthermore, when the permanent magnets were removed, the increase in displacement was no longer observed, indicating that the MNPs continued to function as a working contrast agent, and likely ruling out any confounding time-dependent cause for the increased signal. Crucially, the addition of permanent magnets alone would not be sufficient to allow MRAS measurements up to 200 Hz without significantly degraded forces. Instead, the current system would need to be modified to incorporate solenoids with fewer turns, and thus less inductive reactance. The resulting decrease in force could then be compensated for via the addition of permanent magnets as predicted.

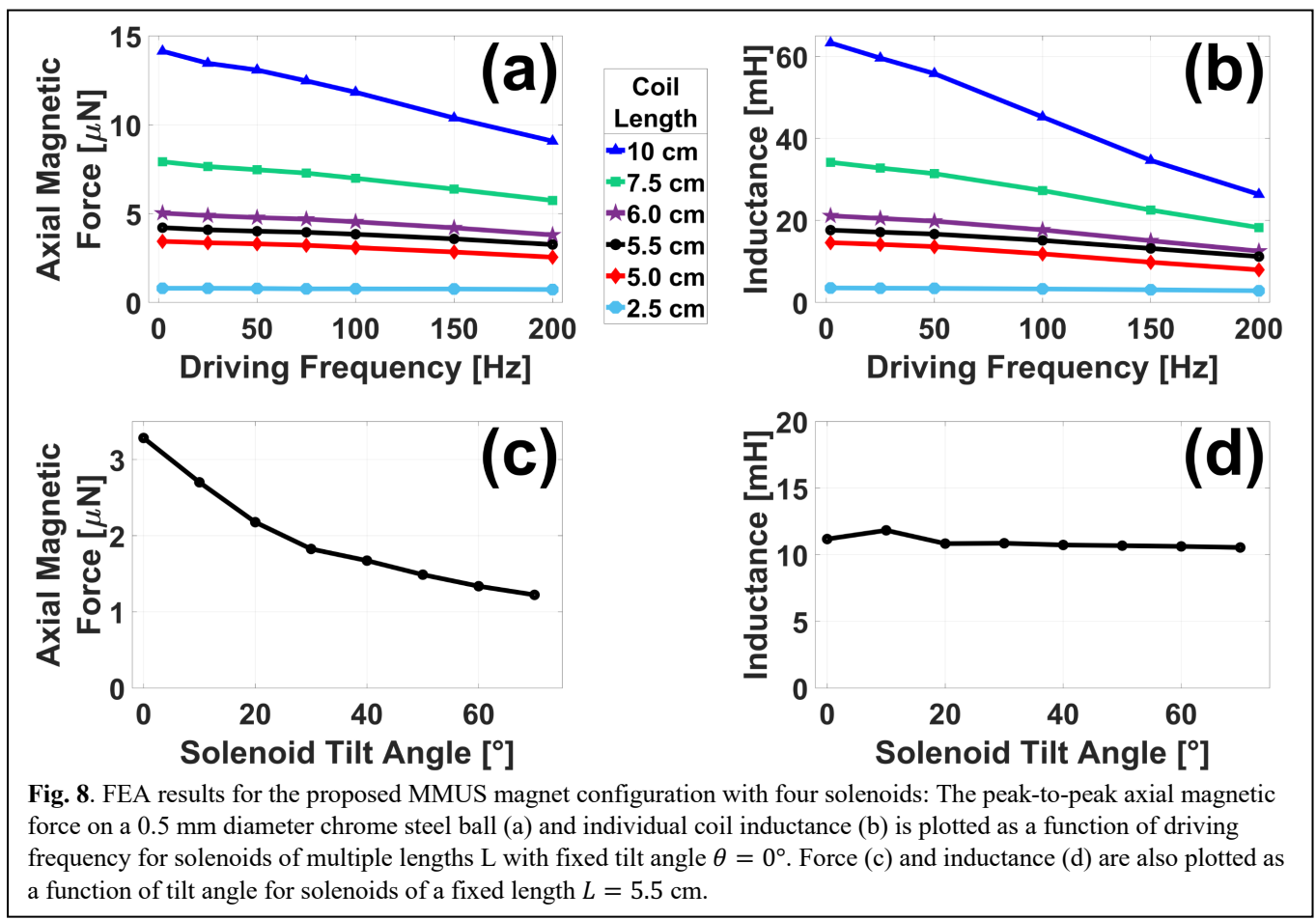

**Fig. 8**. FEA results for the proposed MMUS magnet configuration with four solenoids: The peak-to-peak axial magnetic force on a 0.5 mm diameter chrome steel ball (a) and individual coil inductance (b) is plotted as a function of driving frequency for solenoids of multiple lengths L with fixed tilt angle $\theta = 0°$. Force (c) and inductance (d) are also plotted as a function of tilt angle for solenoids of a fixed length $L = 5.5$ cm.

To address the point that strong permanent magnets could bring additional safety and practicality concerns to MRAS, the FEA model was used to assess a second magnet configuration composed only of solenoids. The model predicted that four solenoid coils, each 5.5 cm in length, wrapped around iron cores 2.0 cm in diameter and 10.5 cm in length could together generate a force comparable to that produced by the current system while maintaining the open-air configuration. Because each coil would only need approximately 384 turns of 18 AWG wire rather than the 1,100 required for the current solenoids, the model predicted this configuration to exhibit significantly less inductance and thus sufficient force up to 200 Hz.

The model presented in this work may have value not only for MMUS elastometry, but also for the MMUS community more broadly where new clinical and pre-clinical applications warrant further study of novel magnet geometries. For example, in magnetic drug targeting and magnetic hyperthermia, MNPs are guided directly to tumors via a magnetic field, allowing for targeted oncological treatments with fewer side effects[36,37]. In order to develop these techniques and to ensure therapies are correctly targeted, imaging modalities capable of localizing MNPs are necessary. While MRI and MPI are commonly used today, MMUS could allow for more rapid, cost-effective localization and concentration measurements with fewer logistical challenges[16,38]. Such applications may require the design of magnet systems specifically tailored to avoid interference with magnetic fields being used for guidance and hyperthermia. The FEA model presented here may be especially useful in this regard.

Long computation times were a key limitation in this work, restricting the feasibility of larger parametric sweeps or optimization studies. For example, each data point in Fig. 6 required an average of 36 min, while each data point in Fig. 8a required an average of 14 minutes. Simulations were run on Windows 10 PCs with 8-core Intel Xeon E5-1660 v4 processors and 32 GB of RAM. If faster computation times are desired, future work may investigate increased use of parallelization[39], the creation of better-optimized meshes via mesh refinement studies, and faster hardware or cluster computing. The inclusion of permanent magnetic fields was found to significantly slow computation in this study, so a focus on solenoid-only magnet configurations could be advantageous.

It is important to note that the FEA simulations did not consider Joule heating. When current flows through a conductor, charge carriers experience inelastic collisions causing energy to be dissipated[25]. The resistance of copper wire increases with temperature[40], so increased resistance could lead to decreased current within MMUS solenoids. The magnetic force felt by MNPs in the imaging area increases with current, so Joule heating could lead to a weaker magnetic force. Heating is observed in the existing experimental MMUS system, so a water-cooling jacket is used, maintaining a lower temperature in the coils. This may have contributed to the agreement between experimental and simulation results at 2.0 Hz. Incorporating Joule heating into future simulations could allow for the design of an apparatus without the need for a bulky water-cooling system.

Another element ignored in the simulations is the skin-effect, although its influence may be minor in the low frequency limit. The skin-effect is a phenomenon in AC circuits where the current distribution shifts to the outside, or "skin," of a conductor at higher frequencies, effectively reducing its cross-sectional area[41]. The resulting increased resistance could lead to decreased current in the MMUS solenoids and thus a decreased force on MNPs in the imaging area. The skin-effect is unlikely to be significant at frequencies up to 200 Hz[25], but further study is warranted to verify this.

Additionally, the model predicts the force on a small chrome steel ball rather than over a magnetically labeled region. A change in force on the ball is used as a proxy for the change throughout the imaging area. This was beneficial in allowing direct comparisons to be made with existing experimental results, but it is not a limitation of the model. Updating the simulation to predict magnetically induced forces over a larger, arbitrarily shaped region would be straightforward. This, however, would likely only be necessary for the study of magnet configurations with very high force heterogeneity over the imaging region. The model may also need to be updated in the future to incorporate the nonlinear magnetization and time dependent properties of MNPs, should the simulation be asked to probe regimes in which these effects dominate.

Finally, as shown in Fig. 7a, the addition of permanent magnets to the existing MMUS system causes the magnetic force to no longer be purely sinusoidal. The frequency and phase locking algorithm still detected the signal as shown in Fig. 7d, though the displacement amplitude could be slightly underestimated as part of the signal may have been lost to higher harmonics. The waveform was predicted to become increasingly non-sinusoidal as magnet strength and proximity to the imaging area increased. Possible solutions include an updated image processing algorithm or the use of multiple solenoids instead of permanent magnets.

## 5. Conclusion

An FEA model designed to predict the performance of MMUS magnets and to assist in the design of future MRAS-based elastometry systems was presented and validated. The COMSOL Multiphysics® files necessary to run each simulation are available online so that they may be adapted for use with other systems[30]. This work was focused primarily on advancing an imaging system that could in the future be used for elastometry of magnetically labeled tissue such as thrombi. Although the specifics of labeling tissue with MNPs was not the focus of this work, it should be noted that the FEA models presented here have the potential to address a key challenge associated with MRAS, and MMUS more generally. It is likely that in a biological setting, the density of iron oxide within a labeled region would be limited, perhaps to 140 fg/platelet or lower[14]. Beyond frequency considerations, these models could allow for the design of MMUS systems capable of producing sufficient magnetically induced displacements for imaging at lower contrast agent concentrations via the maximization of the applied magnetic force.

## Acknowledgements

The authors wish to thank Dr. Amy L. Oldenburg of the University of North Carolina at Chapel Hill's Department of Physics and Astronomy for the use of her equipment, and Mr. Rich Preville of Davidson College's Department of Physics for his technical expertise.

This work was supported in part by the National Institutes of Health, National Heart, Lung, and Blood Institute, under grant R21HL 119928, and by the Department of Defense Air Force Office of Scientific Research under grant FA9550-14-1-0208.